\documentclass[showpacs,preprintnumbers,nofootinbib,prd,superscriptaddress,10pt]{revtex4-1}
\usepackage{graphicx}
\usepackage{xcolor}
\usepackage{lipsum}
\usepackage{amsmath,amssymb}

\begin{document}
\title{On the testability of the Károlyházy model}
\author{Laria Figurato}
\email{laria.figurato@phd.units.it}
\affiliation{Department of Physics, University of Trieste, Strada costiera 11, 34151 Trieste, Italy}
\affiliation{Istituto Nazionale di Fisica Nucleare, Trieste section, Via Valerio 1, 34127 Trieste, Italy}

\author{Angelo Bassi}
\email{bassi@ts.infn.it}
\affiliation{Department of Physics, University of Trieste, Strada costiera 11, 34151 Trieste, Italy}
\affiliation{Istituto Nazionale di Fisica Nucleare, Trieste section, Via Valerio 1, 34127 Trieste, Italy}

\author{Sandro Donadi}
\email{s.donadi@qub.ac.uk}
\affiliation{Centre for Quantum Materials and Technologies, School of Mathematics and Physics, Queen’s University, Belfast BT7 1NN, United Kingdom}

\begin{abstract}
Károlyházy's original proposal, suggesting that space-time fluctuations could be a source of decoherence in space, faced a significant challenge due to an unexpectedly high emission of radiation (13 orders of magnitude more than what was observed in the latest experiment). To address this issue, we reevaluated Károlyházy's assumption that the stochastic metric fluctuation must adhere to a wave equation. By considering more general correlation functions of space-time fluctuations, we resolve the problem and consequently revive the aforementioned proposal.
\end{abstract}

\maketitle

\section{Introduction}

The interplay between Quantum Mechanics and the theory of gravitation is subject to increasing attention, being one of the most important open problems in theoretical physics, from which a far deeper understanding of nature is expected. After decades of intense theoretical research, the goal of a unified theory comprising quantum and gravitational phenomena in a fully consistent way has not yet been reached. From the experimental point of view we know very little about the response of quantum systems to gravity \cite{COW, dieci, undici}, the reason being that it is challenging to reach the regime where masses are big enough for gravitational effects to be detectable, but small enough to prevent environmental noises from compromising their quantum behavior. 

Historically, research on Quantum Gravity focused on the Planck scale, the regime where quantum and gravitational phenomena are equally important \cite{dewitt, ‘t_Hooft, Rovelli,Green}. Due to the theoretical difficulties in unifying the two theories,
 in more recent years attention has grown for assessing the interplay between Quantum Mechanics and the theory of gravitation at low energy, possibly in table-top experiments \cite{Bose,Marletto,Carlesso,anast}. The predicted effects are naturally very small, but in principle observable. 

One of the commonly accepted effects that quantized gravity should have on quantum matter is decoherence \cite{1996, graviton, graviton2, repgravdeco}. This is typical of an open quantum system, in this case, the environment being the gravitational degrees of freedom; once quantized, they fluctuate so that, after tracing them out, the reduced dynamics of the systems show a loss of quantum coherence. Another possibility considered is that gravity is fundamentally classical. In this case, several models involve stochastic spacetime metric \cite{Adlergradeco, giulio, seven} and others describe the gravitational interaction through a continuous
 measurement and feedback protocol \cite{KTM,KTM2,KTM3,TDmodel}. Also in these cases, the modified dynamics induce gravitational decoherence. The magnitude of the effect varies from model to model, and in some cases, experiments have already placed interesting bounds \cite{boundgrav1,boundgrav2,Donadi,no-KTM}.

One of the first proposals for gravitational decoherence was formulated by Károlyházy \cite{K}. He suggested that there is a fundamental limitation in the precision with which a length can be measured using a quantum probe. This limitation, he interpreted as arising due to what he called—with a bit of humor—spacetime haziness, i.e. stochastic fluctuations of the spacetime metric, which prevent measurements from being too sharply defined. A first consequence of these fluctuations is of course decoherence in space, which has been well studied in the literature \cite{K, K_et_al, k_coeff,Diosi3}. Another less known consequence of spacetime haziness is that charged particles, being accelerated by the metric fluctuations, emit radiation; this fact was used to bound the parameters of models predicting the spontaneous collapse of the wave function \cite{fu, Adlerama, Adler, donadler, last, Majo,carlrev}, also excluding some of them \cite{Donadi}.  Diósi and Lukács computed the expected photon emission rate as predicted by the Károlyházy model, showing that it is incompatible with observations \cite{Diosi,Diosi2}. The model is ruled out.

Actually, Károlyházy's original paper resorted to an additional ingredient, besides spacetime fluctuations: it was assumed that these fluctuations have the form of gravitational waves. This assumption constrains the correlation function of the fluctuations, which is ultimately responsible for predicting a large photon emission rate from matter, which in turn is excluded by experimental evidence. Yet, gravitational waves are not necessary to implement  Károlyházy's goal of explaining why there is a fundamental limitation to space measurements; spacetime fluctuations alone, with a more general correlation function, suffice.

In this paper we explore the possibility of generalizing Károlyházy's model by considering a larger class of correlation functions for the gravitational fluctuations, showing that compatibility with current experimental evidence can be restored. 

The paper is organized as follows. In section II we briefly summarize the original model by Károlyházy. In section III we recall the calculation of the radiation emission done by Diósi and Lukács and introduce some basic equations, which will be used later in the paper. In section IV we consider a possible generalization of Károlyházy's model by relaxing the hypothesis that the fluctuations must be of the form of gravitational waves; specifically we consider general correlation functions, which are translational invariant in space and time and symmetric under time reversal. We then find the constraints that the general correlation function must fulfill to be compatible with Károlyházy's idea of the existence of a fundamental limitation to space measurements as well as with experimental data coming from two different experiments. In section V we consider one specific example of a possible correlation function and study its compatibility with experimental data.

\section{A brief review of the Károlyházy model}

The genesis of Károlyházy's model \cite{K,K_et_al} lies in the attempt to answer to the following question: with what precision can a time-like length $s=cT$ in flat spacetime be measured using a quantum probe which obeys the uncertainty principle?
By combining the uncertainty principle together with the requirement that the mass of the system is small enough to guarantee that its Schwarzschild radius is smaller than the uncertainty in the position measurements,  
Károlyházy provided an argument as to why there is a minimum uncertainty $\Delta s$ in the inferred length $s$; the two are related as follows \footnote{In the original work of Károlyházy \cite{K} there is a typo, (which however does not propagate through the rest of the text): in his Eq. (3.1) the power of $s$ is 3/2 while the correct power is 2/3, required for having consistent dimensions on both sides of the equation.} 
\begin{equation}
    \Delta s^3 = l_p^2 s,
    \label{bound}
\end{equation}
with $l_p = \sqrt{\hbar G/c^3}$ being the Planck length \cite{K}.

Károlyházy accounted for this uncertainty by assuming that  spacetime fluctuates around the flat Minkowski metric $\eta_{\mu \nu}$.
The total metric is then given by $(g_{\mu\nu})_\beta (\boldsymbol{x}, t) = \eta_{\mu\nu} (\boldsymbol{x}, t) + (h_{\mu\nu})_\beta (\boldsymbol{x}, t)$,
where $\eta_{\mu\nu} (\boldsymbol{x}, t)$ is the flat Minkowski metric and $(h_{\mu\nu})_\beta (\boldsymbol{x}, t)$ is the ensemble of fluctuating metrics with $\beta$ labelling the different realizations of the metric.
Then, the proper length $s$ between two fixed points in space is defined as the mean of the lengths $s_\beta$ corresponding to the different realizations of the metric
\begin{equation}
    s=\mathbb{E}[s_\beta]=cT,
    \label{s=ct}
\end{equation}
and similarly $\Delta s$ is given by 
\begin{equation}
    \Delta s=\sqrt{\mathbb{E} (s_\beta-s)^2},
    \label{Deltasdef}
\end{equation}
where $\mathbb{E}[...]$ denotes the average over the stochastic metric. 
The relation in Eq. (\ref{bound}) clearly limits the kind of possible stochastic metrics. 

Károlyházy's analysis proceeds by considering non-relativistic dynamics and weak gravitational fields, which imply that only the component $g_{00}$ of the metric is relevant for the dynamics, and that the fluctuations around the Minkowski metric can be written as  
\begin{equation}
    (g_{00})_\beta (\boldsymbol{x}, t) = 1 + \gamma_\beta (\boldsymbol{x}, t),
\end{equation}
where we introduced $\gamma_\beta (\boldsymbol{x}, t)=(h_{00})_\beta (\boldsymbol{x}, t)$.

For each realization of the stochastic metric $\beta$, the line element is given by 
\begin{equation}
    s_\beta = \int_0^T \sqrt{(g_{00})_\beta} c dt = \int_0^T \sqrt{1 + \gamma_\beta (\boldsymbol{x}, t)} c dt \simeq c \int_0^T \Big( 1 + \frac{1}{2} \gamma_\beta (\boldsymbol{x}, t) \Big)  dt.
\end{equation}
In order to obtain Eq. (\ref{s=ct}), it follows that $\mathbb{E}[\gamma_\beta (\boldsymbol{x}, t)]=0$. Moreover, when replaced in Eq. (\ref{Deltasdef}), one gets 
\begin{equation}\label{Delta7}
    \Delta s^2 = \frac{c^2}{4} \mathbb{E}\Big( \int_0^T \gamma_\beta (\boldsymbol{x}, t) dt \Big) ^2 = l_p^{4/3} s^{2/3},
\end{equation}
where the last equality follows by imposing the constraint derived by Károlyházy in Eq. (\ref{bound}). Note that in Eq. (\ref{Delta7}) one considers the correlation functions at the same point in space, i.e. $\boldsymbol{x}= \boldsymbol{x}'$ because Károlyházy defines the proper length as $s=cT$, hence assuming that the probe is not moving in space \cite{K, k_coeff}.

Károlyházy assumed that $\gamma_\beta$ satisfies the wave equation 
\begin{equation}
    \square \gamma_\beta (\boldsymbol{x}, t) = 0;
    \label{box}
\end{equation}
this implies that the Fourier expansion of the $\gamma_\beta$ is of the form  
\begin{equation}
  \gamma_\beta (\boldsymbol{x}, t) = \frac{1}{\sqrt{l^3}} \sum_{\boldsymbol{k}}  \Big[ c_\beta (\boldsymbol k) e^{i(\boldsymbol{k} \cdot \boldsymbol{x} - \omega_{\boldsymbol{k}} t)} + c.c. \Big], 
\label{gamma}
\end{equation}
with $\omega_{\boldsymbol{k}} = |\boldsymbol{k}|c$. Here, in order to work with normalized plane waves, we confined the system in a box with side $l$, which we will later take to infinity.
Károlyházy further assumed that the different modes are independent and that the correlation depends only by a function $F(k)$ of the modulus of $k=|\boldsymbol{k}|$ of the wave vector i.e.
\begin{equation}\label{coeff}
\mathbb{E} [c_\beta(\boldsymbol{k})c_\beta^*(\boldsymbol{k'})]=\delta_{\boldsymbol{k}, \boldsymbol{k'}} F(k). \end{equation}
By replacing the Fourier expansion Eq. (\ref{gamma}) in Eq. (\ref{Delta7}), using Eq. (\ref{coeff}) and taking the continuum limit $l\rightarrow \infty$,
it is straightforward to show that the function $F(k)$ must be\footnote{Compared to the article of Károlyházy \cite{k_coeff}, we have an extra multiplicative factor $\frac{8 \pi^2}{3 \Gamma (\frac{1}{3})}$, which is necessary to satisfy the constraint given in Eq. (\ref{bound}) with the correct constants.}
\begin{equation}\label{FFF}
    F(k) = \frac{8 \pi^2}{3 \Gamma (\frac{1}{3})}\ l_p^{4/3} k^{-5/3}.
\end{equation}
Károlyházy suggested that the model should be considered meaningless for distances smaller than a cutoff $\lambda_c=10^{-15}$ m \cite{K} (which seems reasonable considering that the model was derived in the non-relativistic regime). For the same reason Eq. (\ref{FFF}) is valid only for value of $k\leq 2\pi/\lambda_c$, while for larger values of $k$ one should take $F(k)=0$.

To summarize, in Károlyházy's model matter is subject to spacetime fluctuations and, once quantized, its wave function decoheres. The fluctuations have zero mean, and  correlation function is given by Eq. (\ref{coeff}), with $F(k)$ given by Eq. (\ref{FFF}).

\section{Semiclassical Derivation of the Emission Rate in the Károlyházy model}\label{II}
We present the derivation of the formula for the radiation emission rate for a free particle following the calculation done by Diósi and Lukács in \cite{Diosi}. Many of the equations introduced here will serve as a basis for the analysis carried out in the next sections. 
This derivation is based on a semiclassical approach that, whenever used in connection to other collapse models,  proved to be perfectly consistent with the fully quantum mechanical calculation \cite{Bassi_Donadi, dirk, last}.

The starting point is the Larmor formula relating the total emitted power $P$ to the acceleration $a$ of a charge

\begin{equation}
    P(t) = \frac{e^2}{6 \pi \epsilon_0 c^3} \mathbb{E}[a^2(\boldsymbol{x},t)],
    \label{Lar}
\end{equation}
where $a(\boldsymbol{x},t)$ is the (modulus of the)  acceleration at time $t$ of a particle with  charge $e$, located in $\boldsymbol{x}$. We are interested in computing the radiation emission rate $d \Gamma(t)/d \omega_{\boldsymbol{k}}$, which gives the number of photons that are emitted per unit of time at a given frequency $\omega_{\boldsymbol{k}}$. The total emitted power is related to the radiation emission rate by the relation
\begin{equation}
    P(t) = \int_0^{+\infty} d \omega_{\boldsymbol{k}} \hbar \omega_{\boldsymbol{k}} \frac{d \Gamma(t)}{d \omega_{\boldsymbol{k}}}.
    \label{rel}
\end{equation}
In order to find the emission rate it is convenient to take the Fourier transform of the acceleration
\begin{equation}\label{accFT}  \boldsymbol{a} (\boldsymbol{x}, \omega) = \int_{-\infty}^{+\infty}dte^{-i\omega t}\boldsymbol{a}(\boldsymbol{x},t),
\end{equation}
and insert it into the total radiation power in Eq. (\ref{Lar}), obtaining
\begin{equation}\label{generale}
P(t) = \frac{e^2}{6 \pi \epsilon_0 c^3} \frac{1}{(2 \pi)^2} \sum_{j=1} ^3 \int_{-\infty}^{+\infty} d\nu \int_{-\infty}^{+\infty} d\omega e^{i(\nu + \omega) t}\ \mathbb{E} [ a_j (\boldsymbol{x}, \omega) a_j (\boldsymbol{x}, \nu) ].
\end{equation}

We now specialize the calculation to the Károlyházy model.
The effect of $\gamma_\beta$ perturbing the metric tensor component $g_{00}$
is equivalent to introducing the random potential \cite{Diosi, Bera_Donadi, seven}
\begin{equation}
V_\beta(\boldsymbol{x}, t) = \frac{1}{2} m c^2 \gamma_\beta (\boldsymbol{x}, t)
\label{potenziale}
\end{equation}
into the Schrödinger equation. The corresponding acceleration is
\begin{equation}
    \boldsymbol{a} (\boldsymbol{x}, t)  = - \frac{1}{2} c^2 \nabla \gamma_\beta (\boldsymbol{x}, t), 
  \label{acc}
\end{equation}
and its Fourier transform is (see Eq. (\ref{gamma})) 
\begin{equation}\label{acc}
\boldsymbol{a}(\boldsymbol{x},\omega)=-\frac{c^{2}\pi}{\sqrt{l^{3}}}\sum_{\boldsymbol{k}}i\boldsymbol{k}\left[c(\boldsymbol{k})e^{i\boldsymbol{k}\cdot\boldsymbol{x}}\delta(\omega+\omega_{\boldsymbol{k}})-c^{*}(\boldsymbol{k})e^{-i\boldsymbol{k}\cdot\boldsymbol{x}}\delta(\omega-\omega_{\boldsymbol{k}})\right].
\end{equation}
By using Eq. (\ref{coeff}), the integrand in Eq. (\ref{generale}) is
\begin{equation}
    \mathbb{E}[a_{j}(\boldsymbol{x},\omega)a_{j}(\boldsymbol{x},\nu)]=\frac{c^{4}\pi^{2}}{l^{3}}\left[\sum_{\boldsymbol{k}}k_{j}^{2}F(k)\left[\delta(\omega+\omega_{\boldsymbol{k}})\delta(\nu-\omega_{\boldsymbol{k}})+\delta(\omega-\omega_{\boldsymbol{k}})\delta(\nu+\omega_{\boldsymbol{k}})\right]\right].
\end{equation}
By inserting this in Eq. (\ref{generale}), integrating the Dirac-$\delta$ functions and taking the limit $l\longrightarrow\infty$, one arrives at the following result
\begin{equation}
   P(t)=\frac{e^{2}}{3\pi\epsilon_{0}c^{3}}\frac{1}{(2\pi)^{5}}\int_{-\infty}^{+\infty} d\boldsymbol{k}c^{4}\pi^{2}k^{2}F(k)=\frac{e^{2}c\ell_{P}^{4/3}}{9\pi\epsilon_{0}\Gamma\left(\frac{1}{3}\right)}\int_{0}^{\frac{2\pi}{\lambda_{c}}}dkk^{7/3}=\frac{4\times2^{1/3}\pi^{7/3}e^{2}c\ell_{P}^{4/3}}{15\Gamma\left(\frac{1}{3}\right)\epsilon_{0}\lambda_{c}^{10/3}},
\end{equation}
where in the second step we used Eq. (\ref{FFF}) and integrated over the angular variables. 
The emission rate can be found by comparing this equation to Eq. (\ref{rel}), leading to
\begin{equation}\label{krate}
\frac{d\Gamma(t)}{d\omega_{\boldsymbol{k}}}=\frac{e^{2}\ell_{P}^{4/3}}{9\pi\epsilon_{0}\Gamma\left(\frac{1}{3}\right)c^{7/3}\hbar}\omega_{\boldsymbol{k}}^{4/3}.
\end{equation}
As already discussed in \cite{Diosi}, the Károlyházy model predicts a large amount of radiation emitted due to the spacetime fluctuations, in disagreement with observations. To see this, we compare the predicted emission rate to the data used to set the strongest bounds on the Continuous Spontaneous Localization (CSL) \cite{Pearle,CSL,rep1,report2} and Diósi-Penrose (DP) \cite{Diosi1987,Diosi1989,Penrose} parameters (for a brief introduction to these models see the Appendix A). A direct comparison of Eq. (\ref{krate}) with the data in \cite{Majo} requires a detailed statistical analysis that goes beyond the purpose of this paper; instead, we set an upper bound on the experimental measured emission rate 
 by taking the formula for the emission in the CSL model considered in \cite{Majo} (Eq. (2)) and apply it to the case we are considering of a single particle with charge $e$. The CSL model has two phenomenological parameters: the collapse rate $\lambda$, which sets the strength of the collapse and the correlation length $r_C$ which set the spatial scale of the collapse. The radiation emission rate depends always on the ratio $\lambda/r_C^2$, for details see [28-31].
 
 We take as values for the CSL parameters the ones given by the bound found in the same paper, i.e. $\lambda/r_C^2=(4.94 \pm 0.15)\times 10^{-1} \textrm{s}^{-1} \textrm{m}^{-2}$. Then we get  \begin{equation}\label{ratexp}     \frac{d\Gamma(t)}{d\omega_{\boldsymbol{k}}}\Big|_{\text{\tiny exp}}=\frac{e^{2}\hbar\lambda}{4\pi^{2}\epsilon_{0}m_{0}^{2}c^{3}r_{C}^{2}\omega_{\boldsymbol{k}}}=\frac{5.09\times10^{-35}\;\textrm{s}^{-1}}{\omega_{\boldsymbol{k}}}.
 \end{equation}
In Table \ref{etichetta} we compare the radiation emission rate from a particle with charge $e$, as predicted by the Károlyházy model, with the experimental observed rate found using Eq. (\ref{ratexp}) for three different energy values. In all three cases, Károlyházy's predictions are more than 13 orders of magnitude larger than the observed radiation rate. 
\begin{table}[h!]
    \centering
    \renewcommand\arraystretch{2.7}
    \renewcommand\tabcolsep{6pt}
    \begin{tabular}{|c|c|c|}
         \hline
         \textbf{E }[KeV] & $\mathbf{\frac{d \Gamma(t)}{d \omega_{\boldsymbol{k}}} \Big|_{K}}$ & $\mathbf{\frac{d \Gamma(t)}{d \omega_{\boldsymbol{k}}} \Big|_{exp} }$ \\
         \hline
         1 & $3.7 \cdot 10^{-38}$ & $2.1 \cdot 10^{-52}$ \\
         \hline
         10 & $8.0 \cdot 10^{-37}$ & $2.1 \cdot 10^{-53}$ \\
         \hline
         1000 & $3.7 \cdot 10^{-34}$ & $2.1 \cdot 10^{-55}$ \\
         \hline
  \end{tabular}
  \caption{ Comparison between the radiation emission rate predicted by Károlyházy's model (Eq. (\ref{krate})) and the upper bound on the rate inferred by the experimental data (Eq. (\ref{ratexp})), for three different energy values.
  } \label{etichetta}
\end{table}

\section{Generalization of Károlyházy model and predictions}\label{III}

As discussed in the previous section, the Károlyházy model predicts a large amount of radiation emitted, which is in contrast with observations. At the core of Károlyházy's idea lies the assumption that the stochastic metric should reproduce Eq. (\ref{bound}); however, in formulating his model, Károlyházy further assumes that the metric fluctuations should satisfy a wave equation and therefore be of the form given in Eq. (\ref{gamma}). This leads to a correlation function for the fluctuations $\gamma_\beta$ which is highly non-Markovian and where the spatial and temporal dependencies are not factorized \cite{seven}. 

This behaviour is completely different from what happens in other related models, such as the CSL and DP models, where the spatial and temporal dependencies are factorized and the models predict a much weaker amount of radiation \cite{Donadi, last}. One then might wonder if, by considering factorized correlation functions also in the Károlyházy model, the predicted radiation emission rate might be lower. 
We will answer to this question by tackling the problem in a more general way: we will look for the general form of correlator, which satisfies the constraint in Eq. (\ref{bound}), and then we will consider the experimental consequences. In particular, we will relax the hypothesis that the wave equation in Eq. (\ref{box}) must be fulfilled.

By relaxing the expansion in plane waves, we assume only that the correlation function of the fluctuations $\gamma_\beta (\boldsymbol{x}, t)$ be translational covariant in space and time and symmetric under time reversal, i.e.
\begin{equation}
    \mathbb{E} [ \gamma_\beta (\boldsymbol{x}, t) \gamma_\beta (\boldsymbol{x}', t') ] = g (\boldsymbol{x} - \boldsymbol{x}', |t - t'|).
    \label{correl}
\end{equation}
Using Eq. (\ref{Delta7}), which includes the constraint in Eq. (\ref{bound}), one gets the condition
\begin{equation}
   \int_0^T dt' \int_0^T dt\  g(0, |t-t'|) = 4 \Big(\frac{l_p}{c}\Big)^{\frac{4}{3}} T^{\frac{2}{3}}.
    \label{ds}
\end{equation}
We make the change of variables $\tau=t-t'$ and $\tau_+=t+t'$, perform the integration over $\tau_+$ and then take the second derivative of Eq. (\ref{ds}) with respect to $T$, obtaining \footnote{The fact that the correlation in Eq. (\ref{condiz}) is not a Dirac delta implies that the Károlyházy model is necessarily non-Markovian. This is reflected in the predicted emitted rate in Eq. (\ref{finalrate}), which does not have dependence $1/\omega$ typical of the Markovian models, see for example \cite{Adlerama, donadler}. Note that the non-Markovianity is present also for the original  Károlyházy's proposal (see the master Eq. (6) in \cite{seven}, which is clearly not-Markovian).}
\begin{equation}
    g(0, |T|) =  - \frac{4}{9} \Big(\frac{l_p}{c}\Big)^{\frac{4}{3}} |T|^{-\frac{4}{3}}.
    \label{condiz}
\end{equation}
Let us now introduce the Fourier transform of the correlation function $g(\boldsymbol{y}, |\tau|)$
\begin{equation}
    \tilde{g}(\boldsymbol{k},|\omega|):=\int_{-\infty}^{+\infty} d\boldsymbol{y}\int_{-\infty}^{+\infty} d\tau e^{-i(\boldsymbol{k}\cdot\boldsymbol{y}+\omega\tau)}g(\boldsymbol{y},|\tau|).
    \label{trasf}
\end{equation}
This implies that
\begin{equation}
g(0, |\tau|) = \frac{1}{(2 \pi)^4} \int_{-\infty}^{+\infty} d \omega e^{i \omega \tau} f(|\omega|),
    \label{fou}
\end{equation}
with 
\begin{equation}\label{def}
f(|\omega|):=\int_{-\infty}^{+\infty} d \boldsymbol{k} \Tilde{g}(\boldsymbol{k}, |\omega|). 
\end{equation}
By comparing Eq. (\ref{condiz}) with Eq. (\ref{fou}) we obtain for $f(|\omega|)$ the final relation
\begin{equation}
    f(|\omega|) = (2 \pi)^3 \frac{4 \sqrt{3}}{3} \Big(\frac{l_p}{c}\Big)^{\frac{4}{3}} \Gamma \Big(\frac{2}{3}\Big) |\omega|^{\frac{1}{3}}.
    \label{f}
\end{equation}

To summarize, we showed that given a correlation function of the form $g(\boldsymbol{y}, |\tau|)$, and given its Fourier transform $\tilde{g}(\boldsymbol{k}, |\omega|)$, the Károlyházy condition in Eq. (\ref{bound}) corresponds to the condition in Eq. (\ref{f}) with $f(|\omega|)$ defined in Eq. (\ref{def}).

The next task is to compute the constraints on the correlation function given by experimental data. We will consider two scenarios: radiation emission, as discussed before, and violation of energy conservation, both induced by the gravitational noise. Starting with radiation emission,  given in Eq. (\ref{generale}), which is true in general, we need to compute the correlations
\begin{equation}\label{acc_corr}    
\mathbb{E} [ a_i (\boldsymbol{x}, \omega) a_j (\boldsymbol{x}, \nu) ] = \frac{c^{4}}{4(2\pi)^{2}}\delta(\omega+\nu)\int_{-\infty}^{+\infty} d\boldsymbol{k}k_{i}k_{j}\tilde{g}(\boldsymbol{k},|\omega|).
\end{equation}
Substituting this in Eq. (\ref{generale}) we arrive at the radiation emission rate in terms of $\Tilde{g}(\boldsymbol{k}, |\omega|)$
\begin{equation}\label{pot}
P(t) = \frac{e^2 c}{12 (2 \pi)^5 \epsilon_0} \int_{-\infty}^{+\infty} d \boldsymbol{k} \int_{-\infty}^{+\infty} d \omega k^2 \Tilde{g}(\boldsymbol{k}, |\omega|).
\end{equation}
The emission rate can be found by comparing this equation to Eq. (\ref{rel}), leading to
\begin{equation}\label{finalrate}
    \frac{d \Gamma(t)}{d \omega} = \frac{e^2 c}{6 (2 \pi)^5 \epsilon_0} \frac{1}{\hbar \omega} \int_{-\infty}^{+\infty} d \boldsymbol{k}\ k^2 \Tilde{g}(\boldsymbol{k}, |\omega|),
\end{equation}
where the factor 6 in place of the factor 12 in the denominator is due to the fact that the integral in Eq. (\ref{rel}) goes from 0 to $+\infty$, while the one in Eq. (\ref{pot}) goes from $-\infty$ to $+\infty$. In this way, this is the radiation emission rate for a generic correlation function of the form introduced in Eq. (\ref{correl}). Any function $\Tilde{g}(\boldsymbol{k}, |\omega|)$ is allowed, as long as it satisfies the relation in Eq. (\ref{f}) with $f(|\omega|)$ given by Eq. (\ref{def}) and, at the same time, it predicts a radiation emission rate compatible with  experimental bounds \cite{Majo}. 
Considering the data in \cite{Majo}, which currently sets the strongest bound on the radiation emission rate, one arrives at the bound 
\begin{equation} \label{sdhgfjg}
    \int_{-\infty}^{+\infty} d \boldsymbol{k}\ k^2 \Tilde{g}(\boldsymbol{k}, |\omega|) \le 3.6 \times 10^{-46} \frac{\textrm{s}}{\textrm{m}^2},
\end{equation}
for all $\omega$ corresponding to the energies considered in the experiment, which are in the range of [1-1000] KeV.

We next consider energy violation. All collapse models predict a change of the kinetic energy of a system due to the interaction with the noise inducing the collapse \cite{diffusion}. 
This effect is present also in the Károlyházy model, where the random fluctuations of the metric induce diffusion in momentum. Several experiments based on this energy increase has been considered for the CSL model, ranging from optomechanical devices \cite{vin1,vin2}, to cold atoms \cite{bila}, to thermal emission from planetary systems \cite{carlessonettuno}. Recently, very good bounds have been set by studying the residual heat leak experiments performed
in ultra-low temperature cryostats for both the CSL model \cite{Moh, AV} as well as the DP model \cite{VU}. Here we extend this analysis to the Károlyházy model. The key quantity is the heating rate of the crystal per unit of mass, which we computed in Appendix B:  
\begin{equation}\label{Heating_main}
    \frac{d E(t)}{dMdt}=\frac{c^{4}}{8}\int_{-\infty}^{+\infty}\frac{d\boldsymbol{k}}{(2\pi)^{3}}\tilde{g}(\boldsymbol{k},|k|v_{s})\boldsymbol{k}^{2}
\end{equation}
with $v_{s}$ the speed of sound in the crystal. In deriving Eq. (\ref{Heating_main}) it was assumed that the correlation function $\tilde{g}(\boldsymbol{k},\omega)$ has a cutoff in $\boldsymbol{k}$ at $k=|\boldsymbol{k}|\leq 1/a$ with $a$ the lattice constant which is typically of order $a\sim 10^{-10} - 10^{-9}$ m. Following \cite{VU} and by considering the cryostat described by Gloos et al. \cite{Gloos}, one can safely assume $\frac{d E(t)}{dMdt}\leq 10 \textrm{ pW}/\textrm{Kg}$, which implies 
\begin{equation}\label{condenergy}
    \int_{-\infty}^{+\infty}d\boldsymbol{k}\tilde{g}(\boldsymbol{k},|k|v_{s})\boldsymbol{k}^{2}\leq2.5\times10^{-42}\;\frac{\textrm{s}}{\textrm{m}^2},
\end{equation}
taking into account that $v_{s}=4000$ m/s, which is the speed of sound in copper at low temperatures. 

We conclude this section by writing down the master equation for this model for a system of $N$ point-like particles. In such a case the mass density is given by 
\begin{equation}\label{mu}
\hat{\varrho}(\boldsymbol{x})=\sum_{i=1}^{N}m_{i}\delta(\boldsymbol{x}-\hat{\boldsymbol{q}}_{i}),
\end{equation}
and the potential in Eq. (\ref{potenziale})  generalizes to
\begin{equation}\label{Vmany}
\hat{V}_{\beta}(t)=\frac{c^{2}}{2}\int_{-\infty}^{+\infty} d\boldsymbol{x}\hat{\varrho}(\boldsymbol{x})\gamma_{\beta}(\boldsymbol{x},t).
\end{equation}
The corresponding master equation, which can be found perturbatively with respect to the gravitational fluctuations, is \cite{seven} 
\begin{equation}\label{ME}
\frac{d\hat{\rho}(t)}{dt}=-\frac{i}{\hbar}\left[\hat{H},\hat{\rho}(t)\right]-\left(\frac{c^{2}}{2\hbar}\right)^{2}\int_{-\infty}^{+\infty} d\boldsymbol{x}\int_{-\infty}^{+\infty} d\boldsymbol{x}'\int_{0}^{t}dt'g(\boldsymbol{x}-\boldsymbol{x}',|t-t'|)
\left[\hat\varrho(\boldsymbol{x}),\left[e^{\frac{i}{\hbar}\hat{H}(t'-t)}\hat\varrho(\boldsymbol{x}')e^{-\frac{i}{\hbar}\hat{H}(t'-t)},\hat{\rho}(t)\right]\right].
\end{equation}

\section{A possible choice of the correlation function}\label{V}

In this section we go back to our initial question: is it possible that by considering correlation functions which satisfy the bound in Eq. (\ref{bound}) and are factorized in space and time, like those in the CSL and DP models, we can obtain a radiation emission rate compatible with experimental data? We now show that the answer to the question is positive. 

Let us assume that the correlation function is of the factorized form
\begin{equation}\label{gfact}
g(\boldsymbol{y},|\tau|)= u (\boldsymbol{y}) v(|\tau|),
\end{equation}
then Eq. (\ref{condiz}) becomes
\begin{equation}
    v(|\tau|) = -\frac{4}{9} \Big(\frac{l_p}{c}\Big)^{\frac{4}{3}} \frac{1}{u(0)} |\tau|^{-\frac{4}{3}}
    \label{quaranta}.
\end{equation}

Therefore, a factorized correlation function is compatible with the generalized Károlyházy's model here proposed, as long as Eq. (\ref{quaranta}) is satisfied.

The radiation emission rate  depends on the Fourier transform $\Tilde{g}(\boldsymbol{k}, |\omega|)$ of $g$, which is also factorized, i.e. $\Tilde{g}(\boldsymbol{k}, |\omega|)=\tilde{u}(\boldsymbol{k}) \tilde{v}(|\omega|)$ with 
\begin{equation}
    \tilde{v}(|\omega|)=\int_{-\infty}^{+\infty} d\tau e^{-i\omega\tau}v(|\tau|)=\frac{4\Gamma\left(\frac{2}{3}\right)\left(\frac{\ell_{p}}{c}\right)^{4/3}}{\sqrt{3}u(0)}\left|\omega\right|^{\frac{1}{3}}.
\end{equation}
Then the radiation emission rate in Eq. (\ref{finalrate}) becomes
\begin{equation}
\frac{d\Gamma(t)}{d\omega}=\frac{e^{2}c\tilde{v}(|\omega|)}{6(2\pi)^{5}\epsilon_{0}\hbar\omega}\int_{-\infty}^{+\infty} d\boldsymbol{k}k^{2}\tilde{u}(\boldsymbol{k})=\frac{e^{2}c}{48\pi^{5}\epsilon_{0}\hbar}\frac{\Gamma\left(\frac{2}{3}\right)}{\sqrt{3}}\left(\frac{\ell_{p}}{c}\right)^{4/3}\frac{\int_{-\infty}^{+\infty} d\boldsymbol{k}k^{2}\tilde{u}(\boldsymbol{k})}{u(0)}\omega^{-\frac{2}{3}},
\end{equation}
and the corresponding bound of Eq. (\ref{sdhgfjg}) becomes
\begin{equation}\label{bound46}
\frac{4\Gamma\left(\frac{2}{3}\right)\left(\frac{\ell_{p}}{c}\right)^{4/3}}{\sqrt{3}}\frac{\left|\omega\right|^{\frac{1}{3}}}{u(0)}\int_{-\infty}^{+\infty} d\boldsymbol{k}\ k^{2}\tilde{u}(\boldsymbol{k})\le3.6\times10^{-46}\frac{\textrm{s}}{\textrm{m}^{2}},
\end{equation}
which should be fulfilled for the values of $\omega$ considered in the experiments, which ranges from [$10^{17}-10^{20}$] Hz, corresponding to energies of the emitted photons in the range [1-1000] KeV. This is clearly possible with an appropriate choice of the function $\tilde u$ (for example, is could be a Gaussian).

Regarding the heating in a crystal, the condition in Eq. (\ref{condenergy}) becomes
\begin{equation}\label{condenergy2}
\frac{4\Gamma\left(\frac{2}{3}\right)\left(\frac{\ell_{p}}{c}\right)^{4/3}v_{s}^{1/3}}{\sqrt{3}u(0)}\int_{-\infty}^{+\infty}d\boldsymbol{k}\tilde{u}(\boldsymbol{k})k^{\frac{7}{3}}\leq2.5\times10^{-42}\frac{\textrm{s}}{\textrm{m}^{2}}.
\end{equation}
We now specify the form of the spatial correlation function $u (\boldsymbol{y})$. It is reasonable to expect that fluctuations related to points in space at relatively small distances are larger compared to those between far away points. Keeping this in mind, and inspired by the correlation function of the CSL model \cite{CSL}, we consider the spatial correlation function 
\begin{equation}\label{ugauss}
u(\boldsymbol{y})=e^{-\frac{\boldsymbol{y}^{2}}{R_{K}^{2}}},
\end{equation}
where spatial correlator $R_K$ is a free parameter. 

For this specific choice of spatial correlation function, the bound from radiation  emission in Eq. (\ref{bound46}), considering photons with frequency of $\omega=10^{20}$ Hz (which are the ones providing the strongest bound), implies that 
\begin{equation}\label{Rkrad}
R_{K}\geq 0.11\textrm{ m}.
\end{equation}
On the other hand, the bound in Eq. (\ref{condenergy2}) implies that
\begin{equation}\label{Rknet}
R_{K}\geq 1.8\times10^{-5}\text{ m}.    
\end{equation}

While experiments set lower bounds on $R_K$, one might be also interested in finding an upper bound to this parameter. This can be obtained by requiring that the strength of the  collapse should be enough to guarantee that macroscopic systems localize sufficiently fast. Following the analysis done in \cite{Toros} for the CSL model, we require the collapse to be strong enough to guarantee that a single-layered graphene disk of the size of $10\;\mu$m (roughly the smallest visible size by human eye) is spatially localized in the time $t=0.01$ s (about the smallest time resolution of the human eye). 

In order to compute  the decoherence effects, we start by considering the master equation (\ref{ME}).
We set $\hat{H}=0$, since we are interested in estimating only the spatial  decoherence, and we write the master equation for the matrix element $\hat{\rho}(\boldsymbol{a},\boldsymbol{b},t):=\langle\boldsymbol{a}|\hat{\rho}(t)|\boldsymbol{b}\rangle$, where $|\boldsymbol{a}\rangle:=|\boldsymbol{a}_{1},...,\boldsymbol{a}_{N}\rangle$ and $|\boldsymbol{b}\rangle:=|\boldsymbol{b}_{1},...,\boldsymbol{b}_{N}\rangle$
\begin{equation}
\frac{d\hat{\rho}(\boldsymbol{a},\boldsymbol{b},t)}{dt}\!=\!-\!\left(\frac{c^{2}}{2\hbar}\right)^{2}\!\!\!\int_{0}^{t}\!dt'\!\sum_{i,j}m_{i}m_{j}\left[g(\boldsymbol{a}_{i}-\boldsymbol{a}_{j},t-t')\!-\!g(\boldsymbol{a}_{i}-\boldsymbol{b}_{j},t-t')\!-\!g(\boldsymbol{b}_{i}-\boldsymbol{a}_{j},t-t')\!+\!g(\boldsymbol{b}_{i}-\boldsymbol{b}_{j},t-t')\right]\!\hat{\rho}(\boldsymbol{a},\boldsymbol{b},t).
\end{equation}
If we now specialize to a factorized correlation function of the form in Eq. (\ref{gfact}), the above master equation becomes
\begin{equation}\label{MEcol}
\frac{d\hat{\rho}(\boldsymbol{a},\boldsymbol{b},t)}{dt}=\left[-\left(\frac{c^{2}}{2\hbar}\right)^{2}\left(\int_{0}^{t}dt'v(|t-t'|)\right)F(\boldsymbol{a},\boldsymbol{b})\right]\hat{\rho}(\boldsymbol{a},\boldsymbol{b},t),
\end{equation}
where
\begin{equation}\label{F}
F(\boldsymbol{a},\boldsymbol{b}):=\sum_{i,j}m_{i}m_{j}\left(u(\boldsymbol{a}_{i}-\boldsymbol{a}_{j})-u(\boldsymbol{a}_{i}-\boldsymbol{b}_{j})-u(\boldsymbol{b}_{i}-\boldsymbol{a}_{j})+u(\boldsymbol{b}_{i}-\boldsymbol{b}_{j})\right).\end{equation}
The solution of Eq. (\ref{MEcol}) is
\begin{equation}\label{MEsol}
\hat{\rho}(\boldsymbol{a},\boldsymbol{b},t)=\hat{\rho}(\boldsymbol{a},\boldsymbol{b},0)\exp\left[-\left(\frac{c^{2}}{2\hbar}\right)^{2}\left(\int_{0}^{t}dt'\int_{0}^{t'}dt''v(|t'-t''|)\right)F(\boldsymbol{a},\boldsymbol{b})\right]\hat{\rho}(\boldsymbol{a},\boldsymbol{b},t),
\end{equation}
with the time integrals giving the factor
\begin{equation}
\int_{0}^{t}dt'\int_{0}^{t'}dt''v(|t'-t''|)=\frac{\sqrt{3}\Gamma\left(\frac{1}{3}\right)\Gamma\left(\frac{2}{3}\right)\left(\frac{\ell_{p}}{c}\right)^{4/3}}{\pi u(0)}t^{2/3}.
\end{equation}
To compute $F(\boldsymbol{a},\boldsymbol{b})$, we consider the  spatial  correlation function in Eq. (\ref{ugauss}). Keeping into account the bound in Eq. (\ref{Rkrad}), and that we are considering a disk with radius $10\;\mu$m and a superposition distance $d=10\;\mu$m, we have that $R_K\geq 0.11\; \textrm{m}\gg |\boldsymbol{a}_i-\boldsymbol{b}_j|$. Then all Gaussian correlations can be Taylor expanded with respect to the exponent, obtaining 
\begin{align}\label{Fab}
F(\boldsymbol{a},\boldsymbol{b})&\simeq-\frac{1}{R^{2}}\sum_{i,j}m_{i}m_{j}\left((\boldsymbol{a}_{i}-\boldsymbol{a}_{j})^{2}-(\boldsymbol{a}_{i}-\boldsymbol{b}_{j})^{2}-(\boldsymbol{b}_{i}-\boldsymbol{a}_{j})^{2}+(\boldsymbol{b}_{i}-\boldsymbol{b}_{j})^{2}\right)=\nonumber\\
&=\frac{2}{R^{2}}\sum_{i,j}m_{i}m_{j}(\boldsymbol{a}_{i}-\boldsymbol{b}_{i})\cdot(\boldsymbol{a}_{j}-\boldsymbol{b}_{j}).
\end{align}
Taking into consideration that the disk is a rigid body and the superposition distance is $\boldsymbol{d}$, we have $\boldsymbol{b}_{i}=\boldsymbol{a}_{i}+\boldsymbol{d}$. Then Eq. (\ref{Fab}) further simplifies to
\begin{equation}\label{finalF}
F(\boldsymbol{a},\boldsymbol{b})=\frac{2M^{2}d^{2}}{R_{K}^{2}},
\end{equation}
with $M=2.4 \times 10^{-16}$ Kg the mass of the graphene disk. 

By replacing these results in Eq. (\ref{MEsol}) and requiring the absolute value of exponent to be larger than 10, so that the off diagonal elements are negligible, we obtain
\begin{equation}\label{Rtheo}
R_{K}\leq1.98\;\textrm{m}.
\end{equation}
In summary, we proved that by requiring the collapse to be compatible with experimental data and also being effective in localizing macroscopic systems, the range of values allowed for the parameter $R_K$ is
\begin{equation}\label{finalR}
0.11\;\textrm{m}\leq R_{K}\leq1.98\;\textrm{m},
\end{equation}
i.e. there is about one order of magnitude room for the parameter $R_{K}$ to make the collapse strong enough, while being compatible with experimental data.

The above analysis shows that, by relaxing the requirement that the fluctuation of the spacetime metric $\gamma_\beta$ fulfills a wave equation, it is possible to find models which comply with Károlyházy's condition in Eq. (\ref{bound}) and are, at the same time, compatible with experimental observations.

\section{Conclusions}\label{Conc}

In this paper we reconsidered a proposal due to Károlyházy on spacetime fluctuations as a possible source of decoherence in space. The original proposal of Károlyházy was ruled out because it  predicts a large amount of radiation emission rate from charged particles interacting with the stochastic gravitational background. 

We considered generalizations of the Károlyházy model, where the constrain in Eq. (\ref{bound}) is still fulfilled, while the assumption that the stochastic metrics satisfies a wave equation as in Eq. (\ref{box}) is relaxed. 
Our analysis shows that any function of the form $g (\boldsymbol{x} - \boldsymbol{x}', |t - t'|)$ which satisfies \\
(i) the Károlyházy's condition 
\begin{equation}
\int_{-\infty}^{+\infty} d \boldsymbol{k} \Tilde{g}(\boldsymbol{k}, |\omega|)= (2 \pi)^3 \frac{4 \sqrt{3}}{3} \Big(\frac{l_p}{c}\Big)^{\frac{4}{3}} |\omega|^{\frac{1}{3}} \Gamma \Big(\frac{2}{3}\Big);
\end{equation}
(ii) the experimental radiation emission bound \begin{equation}        \int_{-\infty}^{+\infty} d \boldsymbol{k}\ k^2 \Tilde{g}(\boldsymbol{k}, |\omega|) \le 3.6 \times 10^{-46} \frac{\textrm{s}}{\textrm{m}^2},
\end{equation}
for the values of $\omega$ considered in the experiments, which range from [$10^{17}-10^{20}$] Hz;

(iii) the experimental bound based on the heating of a crystal \begin{equation}\label{58}
 \int_{-\infty}^{+\infty}d\boldsymbol{k}\tilde{g}(\boldsymbol{k},|k|v_{s})\boldsymbol{k}^{2}\leq2.5\times10^{-42}\;\frac{\textrm{s}}{\textrm{m}^2},
\end{equation}
where $v_s=4000$ m/s is the speed of sound in copper at low temperatures, is a valid correlation function $\tilde{g}(\boldsymbol{k},\omega)$, with a cutoff at $|\boldsymbol{k}|\leq k_{max}=10^9$ m$^{-1}$.

In section \ref{V} we provided one example of correlation function which fulfills these conditions. In particular, for a factorized correlation function of the form $g(\boldsymbol{y},|\tau|)= u (\boldsymbol{y}) v(|\tau|)$, with $u(\boldsymbol{y})=e^{-\frac{\boldsymbol{y}^{2}}{R_{K}^{2}}}$ (the form of $v(|\tau|)$  is fixed by the condition set by Károlyházy), we showed that the allowed range of values for the parameter $R_K$ is 
\begin{equation}
0.11\;\textrm{m}\leq R_{K}\leq1.98\;\textrm{m}.    
\end{equation}
This gives precise indications for devising possible experimental tests of the model.



\section{acknowledgments}
The authors acknowledge M. Carlesso, L. Di\'osi and J. L. Gaona-Reyes for useful discussions. L.F. and A.B. acknowledge financial support from the University of Trieste, INFN and the PNRR MUR project PE0000023-NQSTI. A.B. further acknowledges financial support the EIC Pathfinder project QuCoM (GA No. 101046973). 
S.D. acknowledges support from the  Marie Sklodowska Curie Action through the UK Horizon Europe guarantee administered by UKRI, and from INFN.

{\section*{Appendix A: The CSL and DP models}}
We provide a very short summary of the CSL and
DP models and their main equations. More details on the CSL model
can be found in \cite{rep1,report2, Carlesso} and of the DP model in \cite{Diosi,Diosi2,seven}.

In the CSL model, the state vector evolves according to the non-linear
and stochastic equation
\begin{equation}
{\rm d}|\psi_{t}\rangle=\left[-\frac{i}{\hbar}\hat{H}{\rm d}t+\int d\boldsymbol{x}\left(\hat{M}(\boldsymbol{x})-\langle\hat{M}(\boldsymbol{x})\rangle\right){\rm d}W_{t}(\boldsymbol{x})-\frac{1}{2}\int d\boldsymbol{x}\int d\boldsymbol{y}\mathcal{D}(\boldsymbol{x}-\boldsymbol{y})\prod_{\boldsymbol{q}=\boldsymbol{x},\boldsymbol{y}}\left(\hat{M}(\boldsymbol{q})-\langle\hat{M}(\boldsymbol{q})\rangle\right){\rm d}t\right]|\psi_{t}\rangle\label{state CSL}
\end{equation}
where $\hat{H}$ is the Hamiltonian of the system, $\hat{M}(\boldsymbol{x})$
the mass density operator, ${\rm d}W_{t}(\boldsymbol{x})$ a family
of Wiener increments (one for each point of space) with spatial correlation
given by 
\begin{equation}\label{Dcsl}
\mathcal{D}(\boldsymbol{x}-\boldsymbol{y})=\frac{\lambda}{m_{0}^{2}}e^{-\frac{(\boldsymbol{x}-\boldsymbol{y})^{2}}{4r_{C}^{2}}}.
\end{equation}
The first term in Eq. (\ref{state CSL}) gives just the standard quantum
mechanical evolution, while the second and the third terms describe
the wave function collapse. The collapse depends on two phenomenological
parameters: the collapse rate $\lambda$, which sets the strength
of the collapse and the correlation length $r_{C}$. The values of
these parameters are not fixed, but they are constrained by comparison
to experimental data and theoretical requirements. The comparison to experimental data set limits on the strength of the collapse: if in a given experiment we do not observe deviations from quantum mechanics we know the collapse cannot be too strong. On the other hand, we require the collapse to be strong enough to at least guarantee that macroscopic objects in a spatial superposition, collapse very fast, otherwise the model loses its main purpose. These constraints result in bounds for the parameter $\lambda$ and $r_C$, see \cite{Carlesso} for a recent review. One can show that in the allowed range of parameters $\lambda$ and $r_{C}$, the effect
of the last two terms in Eq. (\ref{state CSL}) are negligible for microscopic systems like particles and atoms but, when one consider
a macroscopic system, their effect become dominant and superpositions in space are suppressed. 

Given the stochastic evolution in Eq. (\ref{state CSL}), one can
derive the master equation for the corresponding statistical operator
which is
\begin{equation}
\frac{{\rm d}}{{\rm d}t}\hat{\rho}_{t}=-\frac{\mathrm{i}}{\hslash}\left[\hat{H},\hat{\rho}_{t}\right]+\int{\rm d}\boldsymbol{x}\int{\rm d}\boldsymbol{y}\ \mathcal{D}(\boldsymbol{x}-\boldsymbol{y})[\hat{M}(\boldsymbol{x}),[\hat{M}(\boldsymbol{y}),\hat{\rho}_{t}]].\label{me CSL}
\end{equation}
 Eq. (\ref{me CSL}) is of the Lindblad form, which is a consequence
of the fact that the CSL dynamics is Markovian. 

Regarding the DP model, one can build a state vector equation as well
as the corresponding master equation which have the same structure
of Eq. (\ref{state CSL}) and Eq. (\ref{me CSL}), with the difference
that in this model the noise correlation is given by 
\begin{equation}
\mathcal{D}(\boldsymbol{x}-\boldsymbol{y})=\frac{G}{\hbar}\frac{1}{|\boldsymbol{x}-\boldsymbol{y}|}\label{Ddp},
\end{equation}
and, to avoid divergences,  one is forced to consider smeared out mass densities. The magnitude of this smearing is
given by a length $R_{0}$, which is the only one free parameter of
the model. Currently the strongest bound on the parameter $R_{0}$
comes from experiments related to the radiation emission from matter \cite{Donadi}, which set the lower bound $R_{0}\geq (4.94 \pm 0.15)\times 10^{-10}$ m \cite{Majo}.

\section*{Appendix B: Derivation of Eq. (\ref{condiz}) in the main text}

We derive Eq. (\ref{condiz}) starting from Eq. (\ref{ds}) of the main text, which we report here for convenience

\begin{equation}
   \int_0^T dt' \int_0^T dt\  g(0, |t-t'|) = 4 \Big(\frac{l_p}{c}\Big)^{\frac{4}{3}} T^{\frac{2}{3}}.
   \label{prima}
\end{equation}

We rewrite the left hand side by performing the change of variables $\tau=t-t'$ and $\tau_+=t+t'$

\begin{equation}
   \int_0^T dt \int_0^T dt'\  f(t,t') = \frac{1}{2} \int_0^T d\tau \int_\tau^{2T-\tau} d\tau_+ [f(-\tau,\tau_+) + f(\tau,\tau_+)].
\end{equation}

Since in our case $f(t,t') = g(0,|t-t'|) = g(0,|\tau|)$ we get

\begin{equation}
   \int_0^T dt \int_0^T dt'\  g(0, |t-t'|) = \frac{1}{2} \int_0^T d\tau \int_\tau^{2T-\tau} d\tau_+ [g(0,|\tau|) + g(0,|\tau|)] =
   \nonumber
\end{equation}

\begin{equation}
    = \int_0^T d\tau \int_\tau^{2T-\tau} d\tau_+ g(0,\tau) =  2 \Big[ \int_0^T T g(0,\tau) d\tau - \int_0^T \tau g(0,\tau) d\tau \Big]
\end{equation}

where we removed the absolute values since $\tau$ is always positive within the interval of integration.

Considering again Eq. (\ref{prima}) we derive both sides with respect to $T$

\begin{equation}
     \frac{d}{dT} \left\{ 2 \Big[ \int_0^T T g(0,\tau) d\tau - \int_0^T \tau g(0,\tau) d\tau \Big] \right\} = \frac{d}{dT} \Big[ 4 \Big(\frac{l_p}{c}\Big)^{\frac{4}{3}} T^{\frac{2}{3}} \Big],
\end{equation}

so we have

\begin{equation}
    2 \int_0^T g(0,\tau) d\tau = \frac{8}{3} \Big(\frac{l_p}{c}\Big)^{\frac{4}{3}} T^{-\frac{1}{3}},
\end{equation}

from which

\begin{equation}
     \int_0^T g(0,\tau) d\tau = \frac{4}{3} \Big(\frac{l_p}{c}\Big)^{\frac{4}{3}} T^{-\frac{1}{3}}.
\end{equation}

We derive one more time, obtaining

\begin{equation}
     \frac{d}{dT} \Big[ \int_0^T g(0,\tau) d\tau \Big] = \frac{d}{dT} \Big[ \frac{4}{3} \Big(\frac{l_p}{c}\Big)^{\frac{4}{3}} T^{-\frac{1}{3}} \Big]
\end{equation}

from which finally we get

\begin{equation}
    g(0,T) = -\frac{4}{9} \Big(\frac{l_p}{c}\Big)^{\frac{4}{3}} T^{-\frac{4}{3}}.
\end{equation}

This completes the proof.

\section*{Appendix C: Calculation of the heating rate of a crystal}

\subsection{The random unitary unravelling}

We derive Eq. (\ref{Heating_main}) of the main text, describing
the heating rate per unit of mass of a crystal in the Károlyházy model.

The starting point
is the master equation (\ref{ME}) that we report for convenience
\begin{equation}
\frac{d\hat{\rho}(t)}{dt}=-\frac{i}{\hbar}\left[\hat{H},\hat{\rho}(t)\right]-\left(\frac{c^{2}}{2\hbar}\right)^{2}\int_{-\infty}^{+\infty}d\boldsymbol{x}\int_{-\infty}^{+\infty}d\boldsymbol{x}'\int_{0}^{t}dt'g(\boldsymbol{x}-\boldsymbol{x}',t-t')\left[\hat{\varrho}(\boldsymbol{x}),\left[e^{\frac{i}{\hbar}\hat{H}(t'-t)}\hat{\varrho}(\boldsymbol{x}')e^{-\frac{i}{\hbar}\hat{H}(t'-t)},\hat{\rho}(t)\right]\right]\label{me_Mo},
\end{equation}
where we relaxed the
requirement that the correlation function must depend on $|t-t'|$ and we consider a mass density for the particles
which is not necessarily point-like i.e. 
\begin{equation}
\hat{\varrho}(\boldsymbol{x})=\sum_{i}m_{i}\mu(\boldsymbol{x}-\hat{\boldsymbol{q}}_{i}),
\end{equation}
with $m_i$ the mass of the $i$-th particle here and $\mu(\boldsymbol{x}-\hat{\boldsymbol{q}}_i)$ describes the shape of the mass density. These generalizations do not affect the following derivation. 

We derive the heating rate of a crystal following closely the derivation done in \cite{Moh} for the CSL model.
An alternative derivation, which resort to a perturbative approach,
can be found in \cite{AV}. 

We start by rewriting the master equation in a form closer to Eq.
(7) in \cite{Moh}. To do this we Fourier expand the mass density operator
obtaining
\begin{equation}
\hat{\varrho}(\boldsymbol{x})=\frac{1}{(2\pi)^{3}}\int d\boldsymbol{k}\sum_{i}m_{i}\tilde{\mu}(\boldsymbol{k})e^{i\boldsymbol{k}\cdot(\boldsymbol{x}-\hat{\boldsymbol{q}}_{i})}=\frac{m_{0}}{(2\pi)^{3}}\int d\boldsymbol{k}e^{i\boldsymbol{k}\cdot\boldsymbol{x}}\tilde{\mu}(\boldsymbol{k})\hat{L}(\boldsymbol{k})\label{rhoMo},
\end{equation}
where we used the definitions for the Fourier and the anti-Fourier
transforms
\begin{equation}
\tilde{\mu}(\boldsymbol{k}):=\int d\boldsymbol{x}\mu(\boldsymbol{x})e^{-i\boldsymbol{k}\cdot\boldsymbol{x}}\label{muF_Mo}
\end{equation}
\begin{equation}
\mu(\boldsymbol{x})=\frac{1}{(2\pi)^{3}}\int d\boldsymbol{k}\tilde{\mu}(\boldsymbol{k})e^{i\boldsymbol{k}\cdot\boldsymbol{x}},
\end{equation}
and where
\begin{equation}
\hat{L}(\boldsymbol{k}):=\sum_{i}\frac{m_{i}}{m_{0}}e^{-i\boldsymbol{k}\cdot\hat{\boldsymbol{q}}_{i}}.
\end{equation}
By replacing Eq. (\ref{rhoMo}) in the Eq. (\ref{me_Mo}) and carrying
on the integrals one finally gets 
\begin{equation}
\frac{d\hat{\rho}(t)}{dt}=-\frac{i}{\hbar}\left[\hat{H},\hat{\rho}(t)\right]-\left(\frac{c^{2}}{2\hbar}\right)^{2}\frac{m_{0}^{2}}{(2\pi)^{3}}\int d\boldsymbol{k}\int_{0}^{t}dt'\tilde{h}(\boldsymbol{k},t-t')\left[\hat{L}^{\dagger}(\boldsymbol{k}),\left[e^{\frac{i}{\hbar}\hat{H}(t'-t)}\hat{L}(\boldsymbol{k})e^{-\frac{i}{\hbar}\hat{H}(t'-t)},\hat{\rho}(t)\right]\right],
\end{equation}
where 
\begin{equation}
\tilde{h}(\boldsymbol{k},t-t'):=|\tilde{\mu}(\boldsymbol{k})|^{2}\tilde{g}(\boldsymbol{k},t-t'),
\end{equation}
with $\tilde{\mu}(\boldsymbol{k})$ defined in Eq. (\ref{muF_Mo})
and
\begin{equation}
\tilde{g}(\boldsymbol{k},t-t')=\int d\boldsymbol{x}g(\boldsymbol{x},t-t')e^{-i\boldsymbol{k}\cdot\boldsymbol{x}}.
\end{equation}

Given this master equation, one can find a corresponding unitary unravelling, which is given by
a stochastic Schr\"odinger equation of the form 
\begin{equation}
i\hbar\frac{d|\psi(t)\rangle}{dt}=\left[\hat{H}+\hat{V}(t)\right]|\psi(t)\rangle,
\end{equation}
where 
\begin{equation}
\hat{V}(t)=-\hbar\int\frac{d\boldsymbol{k}}{(2\pi)^{3}}w(t,\boldsymbol{k})\hat{L}(\boldsymbol{k})\label{V_MO},
\end{equation}
with $w(t,\boldsymbol{k})$ a noise with zero average and correlation
\begin{equation}
\mathbb{E}\left[w(t,\boldsymbol{k})w(t',\boldsymbol{k}')\right]=(2\pi)^{3}\frac{m_{0}^{2}c^{4}}{4\hbar^{2}}\tilde{h}(-\boldsymbol{k},t-t')\delta(\boldsymbol{k}+\boldsymbol{k}').\label{corr_MO}
\end{equation}
Note that the requirement that $\hat{V}(t)$ is hermitian implies
$w(t,\boldsymbol{k})=w^{*}(t,-\boldsymbol{k}).$

Eq. (\ref{V_MO}) should be compared to the potential in Eq. (4) of \cite{Moh}. We see that the two models are identical as long as one
replaces 
\begin{equation}
e^{-r_{C}^{2}\boldsymbol{k}^{2}}\xi(t,\boldsymbol{k})\longrightarrow w(t,\boldsymbol{k})\label{corresp_Mo},
\end{equation}
with the correlation of the $w(t,\boldsymbol{k})$ given in Eq. (\ref{corr_MO}).\  

\subsection{Calculation of the heating rate}

Given a crystal, the total energy at time $t$ is given by
\begin{equation}
\mathbb{E}\left[H(t)\right]=\sum_{\boldsymbol{k}\in BZ}\sum_{s=1}^{3r}\hbar\omega_{\boldsymbol{k}s}\mathbb{E}\left[\hat{a}_{\boldsymbol{k}s}^{\dagger}(t)\hat{a}_{\boldsymbol{k}s}(t)\right]\label{H_Mo},
\end{equation}
where the first sum is over the mode in the first Brillouin zone ($BZ$), the second sum is over the acoustic and optical branches and $\hat{a}_{\boldsymbol{k}s}(t)$ and $\hat{a}^{\dagger}_{\boldsymbol{k}s}(t)$
are, respectively, the phonons annihilation and creation operators evolved in the Heisenberg
picture at time $t$. Having established the connection in Eq. (\ref{corresp_Mo})
between our model and the CSL model studied in \cite{Moh}, we can directly
take the solution for the operator in Eq. (21) of \cite{Moh} and
use the replacement in Eq. (\ref{corresp_Mo}). Then, using the notation
of \cite{Moh}, which we also refer to the reader for a full explanation
of all symbols, we get
\begin{equation}
\hat{a}_{\boldsymbol{k}s}(t)=e^{-i\omega_{\boldsymbol{k}s}t}\hat{a}_{\boldsymbol{k}s}+i\int\frac{d\tilde{\boldsymbol{k}}}{(2\pi)^{3}}\int_{0}^{t}dt'e^{-i\omega_{\boldsymbol{k}s}(t-t')}w(\tilde{\boldsymbol{k}},t')\sum_{i,\nu}\frac{m_{\nu}}{m_{0}}e^{-i\tilde{\boldsymbol{k}}\cdot(\boldsymbol{R}_{i}+\boldsymbol{d}_{\nu})}\eta_{i\nu;\boldsymbol{k}s}(\tilde{\boldsymbol{k}})\label{aks_Mo},
\end{equation}
where 
\begin{equation}
\eta_{i\nu;\boldsymbol{k}s}(\tilde{\boldsymbol{k}})=-i\left(\frac{\hbar}{2Nm_{\nu}\omega_{\boldsymbol{k}s}}\right)^{\frac{1}{2}}\sum_{\alpha}\tilde{k}_{\alpha}\boldsymbol{\epsilon}_{\alpha,\nu}^{(s)*}(\boldsymbol{k})e^{-i\boldsymbol{k}\cdot\boldsymbol{R}_{i}}\label{eta_Mo}.
\end{equation}
Inserting Eq. (\ref{aks_Mo}) in Eq. (\ref{H_Mo}), using Eq. (\ref{corr_MO})
and carrying out the calculation we obtain 
\[
\mathbb{E}\left[\hat{H}(t)\right]=\hat{H}+\frac{m_{0}^{2}c^{4}}{4\hbar^{2}}\sum_{\boldsymbol{k}\in BZ}\sum_{s=1}^{3r}\hbar\omega_{\boldsymbol{k}s}\int\frac{d\tilde{\boldsymbol{k}}}{(2\pi)^{3}}\int_{0}^{t}dt'\int_{0}^{t}dt'e^{i\omega_{\boldsymbol{k}s}(t'-t'')}\tilde{h}(\tilde{\boldsymbol{k}},t''-t')\times
\]
\begin{equation}
\times\left(\sum_{j,\kappa}\frac{m_{\kappa}}{m_{0}}e^{i\tilde{\boldsymbol{k}}\cdot(\boldsymbol{R}_{j}+\boldsymbol{d}_{\kappa})}\eta_{j\kappa;\boldsymbol{k}s}^{*}(\tilde{\boldsymbol{k}})\sum_{i,\nu}\frac{m_{\nu}}{m_{0}}e^{-i\tilde{\boldsymbol{k}}\cdot(\boldsymbol{R}_{i}+\boldsymbol{d}_{\nu})}\eta_{i\nu;\boldsymbol{k}s}(\tilde{\boldsymbol{k}})\right).
\end{equation}
By replacing the definition in Eq. (\ref{eta_Mo})
and using $\sum_{\boldsymbol{k}\in BZ}=\frac{N}{V_{BZ}}\int_{BZ}d\boldsymbol{k}$, we get\footnote{See Eq. (1.2.13) in \cite{Callaway},  which is different from the one reported in \cite{Moh} but necessary to obtain the correct result.} 
\[
\mathbb{E}\left[\hat{H}(t)\right]=\hat{H}+\frac{m_{0}^{2}c^{4}}{4\hbar^{2}}\frac{\hbar^{2}}{2V_{BZ}}\sum_{\kappa,\nu}\frac{\sqrt{m_{\kappa}m_{\nu}}}{m_{0}^{2}}\int\frac{d\tilde{\boldsymbol{k}}}{(2\pi)^{3}}e^{i\tilde{\boldsymbol{k}}\cdot(\boldsymbol{d}_{\kappa}-\boldsymbol{d}_{\nu})}\int_{BZ}d\boldsymbol{k}\sum_{i,j}e^{i(\tilde{\boldsymbol{k}}+\boldsymbol{k})\cdot(\boldsymbol{R}_{j}-\boldsymbol{R}_{i})}\times
\]
\begin{equation}
\times\sum_{s=1}^{3r}\int_{0}^{t}dt'\int_{0}^{t}dt'e^{i\omega_{\boldsymbol{k}s}(t'-t'')}\tilde{h}(\tilde{\boldsymbol{k}},t''-t')\left(\sum_{\alpha,\beta}\tilde{k}_{\beta}\tilde{k}_{\alpha}\boldsymbol{\epsilon}_{\alpha,\nu}^{(s)*}(\boldsymbol{k})\boldsymbol{\epsilon}_{\beta,\kappa}^{(s)}(\boldsymbol{k})\right),
\end{equation}
which corresponds to Eq. (29) of \cite{Moh}. We proceed by using (see
Eqs. (30-32) in \cite{Moh})
\begin{equation}
\int_{0}^{t}dt_{1}\int_{0}^{t}dt_{2}e^{-i\omega_{\boldsymbol{k}s}(t_{2}-t_{1})}\tilde{h}(\tilde{\boldsymbol{k}},t_{2}-t_{1})\approx t\tilde{H}(\tilde{\boldsymbol{k}},\omega_{\boldsymbol{k}s}),
\end{equation}
where we introduced 
\begin{equation}
\tilde{h}(\tilde{\boldsymbol{k}},t)=\frac{1}{2\pi}\int_{-\infty}^{+\infty}d\omega\tilde{H}(\tilde{\boldsymbol{k}},\omega)e^{i\omega t},
\end{equation}
obtaining
\[
\mathbb{E}\left[\hat{H}(t)\right]=\hat{H}+\frac{m_{0}^{2}c^{4}}{4\hbar^{2}}\frac{\hbar^{2}t}{2V_{BZ}}\sum_{\kappa,\nu}\frac{\sqrt{m_{\kappa}m_{\nu}}}{m_{0}^{2}}\int\frac{d\tilde{\boldsymbol{k}}}{(2\pi)^{3}}e^{i\tilde{\boldsymbol{k}}\cdot(\boldsymbol{d}_{\kappa}-\boldsymbol{d}_{\nu})}\int_{BZ}d\boldsymbol{k}\sum_{i,j}e^{i(\tilde{\boldsymbol{k}}+\boldsymbol{k})\cdot(\boldsymbol{R}_{j}-\boldsymbol{R}_{i})}\times
\]
\begin{equation}
\times\sum_{s=1}^{3r}\tilde{H}(\tilde{\boldsymbol{k}},\omega_{\boldsymbol{k}s})\sum_{\alpha,\beta}\tilde{k}_{\beta}\tilde{k}_{\alpha}\boldsymbol{\epsilon}_{\alpha,\nu}^{(s)*}(\boldsymbol{k})\boldsymbol{\epsilon}_{\beta,\kappa}^{(s)}(\boldsymbol{k}).
\end{equation}

We now assume that there is a cutoff in $\tilde{H}(\boldsymbol{k},\omega_{\boldsymbol{k}s})$
with respect to $\boldsymbol{k}$, i.e. that there exists a characteristic length
$R$ such that only the modes with $k\leq1/R$ are relevant. When $R\geq a\sim 10^{-10} - 10^{-9}$
m (the typical lattice distances), all relevant $\boldsymbol{k}$'s
are within the first Brillouin zone and we can use
\begin{equation}
\sum_{i}e^{i(\tilde{\boldsymbol{k}}+\boldsymbol{k})\cdot\boldsymbol{R}_{i}}=V_{BZ}\delta(\tilde{\boldsymbol{k}}+\boldsymbol{k}),
\end{equation}
which implies
\begin{equation}
\sum_{i,j}e^{i(\tilde{\boldsymbol{k}}+\boldsymbol{k})\cdot(\boldsymbol{R}_{j}-\boldsymbol{R}_{i})}=V_{BZ}\delta(\tilde{\boldsymbol{k}}+\boldsymbol{k})N.
\end{equation}
Then we get 
\[
\mathbb{E}\left[\hat{H}(t)\right]=\hat{H}+\frac{m_{0}^{2}c^{4}}{4\hbar^{2}}\frac{\hbar^{2}tN}{2}\frac{1}{m_{0}^{2}}\int\frac{d\boldsymbol{k}}{(2\pi)^{3}}\sum_{s=1}^{3r}\tilde{H}(-\boldsymbol{k},\omega_{\boldsymbol{k}s})\left|\sum_{\kappa}\sum_{\beta}\sqrt{m_{\kappa}}k_{\beta}\boldsymbol{\epsilon}_{\beta,\kappa}^{(s)}(\boldsymbol{k})e^{-i\boldsymbol{k}\cdot\boldsymbol{d}_{\kappa}}\right|^{2}.
\]
Again because of the cutoff in $\boldsymbol{k}$, we can approximate $e^{-i\boldsymbol{k}\cdot\boldsymbol{d}_{\kappa}}\simeq1$
and $\boldsymbol{\epsilon}_{\beta,\kappa}^{(s)}(\boldsymbol{k})\simeq\boldsymbol{\epsilon}_{\beta,\kappa}^{(s)}(0)$
and following \cite{Moh} one can see that the only relevant contribution
comes from the acoustic longitudinal branch, which simplifies the formula
to 
\begin{equation}
\mathbb{E}\left[\hat{H}(t)\right]=\hat{H}+\frac{c^{4}}{8}Mt\int\frac{d\boldsymbol{k}}{(2\pi)^{3}}\tilde{H}(\boldsymbol{k},\omega_{\boldsymbol{k}LA})\boldsymbol{k}^{2},\label{afinal_ME}
\end{equation}
where $M=N\sum_{\kappa}m_{\kappa}$ is the total mass of the crystal,
$\omega_{\boldsymbol{k}LA}=|k|v_{s}$ with $v_{s}$ the velocity of
sound in the crystal and we performed the change of variable $\boldsymbol{k}\rightarrow-\boldsymbol{k}$.
This equation is the analog of Eq. (38) of \cite{Moh}.

By considering the quantum expectation value over the initial state $E(t):=\langle\psi_{i}|\mathbb{E}\left[\hat{H}(t)\right]|\psi_{i}\rangle$
and the heating per unit of mass and time we arrive at 
\begin{equation}
\frac{dE(t)}{dMdt}=\frac{c^{4}}{8}\int_{-\infty}^{+\infty}\frac{d\boldsymbol{k}}{(2\pi)^{3}}\tilde{H}(\boldsymbol{k},|k|v_{s})\boldsymbol{k}^{2}.\label{finalME}
\end{equation}
Note that Eq. (88) is completely independent from the initial
state $|\psi_{i}\rangle$ of the crystal.

In general, given the definitions above, one can see that 
\begin{equation}
\tilde{H}({\boldsymbol{k}},\omega)=|\tilde{\mu}(\boldsymbol{k})|^{2}\tilde{g}(\boldsymbol{k},\omega).
\end{equation}
For point-like mass densities, which are those we are interested in, $\tilde{\mu}(\boldsymbol{k})=1$  and $\tilde{H}({\boldsymbol{k}},\omega)=\tilde{g}(\boldsymbol{k},\omega)$; In this case Eq. (\ref{finalME}) reduces to Eq. (\ref{Heating_main}) in the main text.

\bibliographystyle{apsrev4-1}

\end{document}